**Stripe phases in $WSe_2/WS_2$ moiré superlattices**

Chenhao Jin[1†*], Zui Tao[1,2†], Tingxin Li[2†], Yang Xu[2], Yanhao Tang[2], Jiacheng Zhu[2], Song Liu[3], Kenji Watanabe[4], Takashi Taniguchi[4], James C. Hone[3], Liang Fu[5*], Jie Shan[1,2,6*], Kin Fai Mak[1,2,6*]

[1] Kavli Institute at Cornell for Nanoscale Science, Ithaca, NY, USA

[2] School of Applied and Engineering Physics, Cornell University, Ithaca, NY, USA

[3] Department of Mechanical Engineering, Columbia University, New York, NY, USA.

[4] National Institute for Materials Science, 1-1 Namiki, Tsukuba, Japan.

[5] Department of Physics, Massachusetts Institute of Technology, Cambridge, MA, USA.

[6] Laboratory of Atomic and Solid State Physics, Cornell University, Ithaca, NY, USA.

† These authors contributed equally to this work

* Correspondence to: jinchenhao@cornell.edu, liangfu@mit.edu, jie.shan@cornell.edu, km627@cornell.edu

**Stripe phases, in which the rotational symmetry of charge density is spontaneously broken, occur in many strongly correlated systems with competing interactions[1-10]. One representative example is the copper-oxide superconductors, where stripe order is thought to be relevant to the mechanism of high-temperature superconductivity[1-3]. Identifying and studying the stripe phases in conventional strongly correlated systems are, however, challenging due to the complexity and limited tunability of these materials. Here we uncover stripe phases in $WSe_2/WS_2$ moiré superlattices with continuously gate-tunable charge densities by combining optical anisotropy and electronic compressibility measurements. We find strong electronic anisotropy over a large doping range peaked at 1/2 filling of the moiré superlattice. The 1/2-state is incompressible and assigned to a (insulating) stripe crystal phase. It can be continuously melted by thermal fluctuations around 35 K. The domain configuration revealed by wide-field imaging shows a preferential alignment along the high-symmetry axes of the moiré superlattice. Away from 1/2 filling, we observe additional stripe crystals at commensurate filling 1/4, 2/5 and 3/5. The anisotropy also extends into the compressible regime of the system at incommensurate fillings, indicating the presence of electronic liquid crystal states. The observed filling-dependent stripe phases agree with the theoretical phase diagram of the extended Hubbard model on a triangular lattice in the flat band limit. Our results demonstrate that two-dimensional semiconductor moiré superlattices are a highly tunable platform to study the stripe phases and their interplay with other symmetry breaking ground states.**

Stripe phases occur in electronic systems generally as a compromise between (effective) short-range attractive interactions and long-range repulsive interactions. They are defined by spontaneous rotational symmetry breaking, and can be classified into stripe crystals and electronic liquid crystals according to insulating or metallic behaviors[1]. Stripe phases have been observed in a variety of correlated systems such as high-temperature superconductors[1-3], quantum Hall systems[4-6], ultracold gases[7], and recently twisted bilayer graphene[8-10]. Formation of stripe phases causes ‘dynamical dimensionality reduction’ and substantially modifies the properties of the system. Such a concept has been invoked to understand the high-temperature superconductivity and the remarkable normal-state properties in the cuprates[1-3]. Due to the competing-interaction origin stripe phases often emerge at the boundary of or coexist with other orders[1-3]. Identifying stripe order and understanding its interplay with other orders remain one of the central challenges in strongly correlated systems.

Two-dimensional (2D) moiré superlattices have recently emerged as a powerful platform to engineer strong correlation physics, such as superconductivity, correlated insulating and Wigner crystal states[8-23]. In particular, semiconducting transition metal dichalcogenide (TMD) moiré superlattices open up new opportunities to detect and study stripe phases. The electronic occupancy can be continuously tuned electrostatically through the entire flat miniband, making it possible to explore filling-tuned quantum melting of stripe crystals and to understand their interplay with other correlated states. The strong light-matter interaction in 2D semiconductors further allows us to directly probe and image stripe order using high-sensitivity optical methods.

Here we report an observation of stripe phases in near-zero-twist-angle $WSe_2/WS_2$ heterostructures, which form a triangular moiré superlattice. In the absence of a stripe phase, the linear electronic response of the system in the 2D plane is isotropic. We probe the electronic anisotropy and compressibility of the heterostructures through resonant optical conductivity and penetration capacitance measurements, respectively, over a wide range of doping density and temperature. Figure 1a and b show the optical microscope image and side-view illustration of a representative device D1. The near 4% lattice mismatch between $WSe_2$ and $WS_2$ gives a moiré periodicity of about 8 nm and a carrier density $n \approx 1.9\times10^{12}/cm^2$ for filling factor $v = 1$, that is, one charge per moiré superlattice cell. Figure 1c shows the gate-dependent optical reflection contrast spectrum at 5 K on the electron-doping side. The dashed box encloses the doping region between $v = 0$ and 1. The result is consistent with previous reports[17,18]. In the absence of doping (gate voltage < 1.3 V), the three peaks near 1.7 eV correspond to the intralayer moiré exciton resonances of $WSe_2$ (Ref. [24-27]). The increase of the resonance amplitude at $v = 1$ (gate voltage 2.25 V) reflects the emergence of a Mott insulating state[17,18] or a charge-transfer insulating state[28] that is originated from a strong on-site Coulomb repulsion. The moiré exciton response has negligible changes over the entire temperature range of study (5 - 40 K) (see supplementary notes 5). Details on the device fabrication and filling factor calibration are provided in Methods.

We focus on the doping region between $v = 0$ and 1, in which strong electronic correlations have been reported[17,18]. Figure 1d shows the filling-dependent penetration capacitance (black), which characterizes how well the heterostructure screens an applied out-of-plane electric field. Each peak corresponds to an incompressible (i.e. insulating) state, and the peak area gives the gap size. We identify the reported Mott or charge-transfer insulating state at $v = 1$ (magenta arrow)[17,18,28] and the two generalized Wigner crystal states at $v = 1/3$ and 2/3 originated from long-range Coulomb interactions (blue arrows)[17]. We also observe other insulating states, including a prominent peak at $v = 1/2$ (red arrow) and several weaker peaks at $v = 1/4$ (purple arrow), 2/5 (green arrow), etc. These are likely also charge-ordered states from long-range Coulomb interactions. In Fig. 1d we also compare the penetration capacitance with interlayer exciton photoluminescence (PL) intensity (blue) (see supplementary notes 4 for the PL spectrum). In the low doping regime below $v = 2/3$, the insulating states show enhanced interlayer exciton PL. This property allows us to use PL, which can be measured under identical experimental conditions as the optical anisotropies, to identify the insulating states locally (spatial resolution is about 800 nm, see Methods for measurements details).

We probe the electronic anisotropy in $WSe_2/WS_2$ moiré superlattices optically. The response is crucially enhanced by choosing a photon energy close to the lowest-energy moiré exciton resonance of $WSe_2$ (see supplementary notes 7). We quantify the electronic anisotropy using order parameter $\eta = \sigma_a - \sigma_b$, where $\sigma_a$ and $\sigma_b$ are the optical conductivity along the anisotropy axis $a$ and $b$, respectively (Fig. 1e). When linearly polarized light (after passing through a linear polarizer P1) impinges on the sample at normal incidence, the electronic anisotropy effectively induces a polarization rotation unless the incident light polarization aligns with one of the anisotropy axes. The rotation angle $\theta$ depends on $\eta$ linearly for $|\eta/\sigma_a| \ll 1$ (see supplementary notes 3 for derivation). We measure $\theta$ from the reflected light intensity $R(\theta)$ by introducing an analyzer (P2) after the sample. We set the analyzer at angle $\phi$ from the cross-polarization configuration. The relative intensity change in the presence of anisotropy, $RC = [R(\theta)/R(\theta = 0) - 1] = [\sin^2(\phi + \theta)/\sin^2\phi - 1] \approx 2\theta/\phi$ (for $\theta \ll \phi$), can be significantly enhanced by choosing a small $\phi$. The quantity $RC$ also changes sign together with $\phi$. These properties allow us to isolate the small anisotropy signal. The anisotropy axes can be further determined by measuring $\theta$ as a function of sample orientation.

Figure 2a shows the doping-dependent $RC$ for one position on device D1 at 5 K with $\phi$ = 0.6° and -0.6° (upper panel). A clear $RC$ signal is observed to peak at $\nu$ = 1/2 (red dashed line) and change sign with $\phi$. Multiple control experiments, including the probe photon energy and power dependences (see supplementary notes 7), verify that the observed anisotropy is an intrinsic response of the doped electrons. The asymmetry between the two measurements is caused by the $\theta^2$-term that is not negligible when $\theta$ is comparable to $\phi$. Meanwhile the PL intensity (lower panel) also shows a prominent enhancement at $\nu$ = 1/2, reflecting its insulating nature. We thus assign the 1/2-state to a stripe crystal. In contrast, the $\nu$ = 1/3 and 2/3 states do not show enhanced anisotropy, although they have similar responses as the 1/2-state in the PL (blue arrows in the lower panel) and penetration capacitance. This is consistent with the previous assignment of the 1/3 and 2/3 states to isotropic electron crystals[17]. These results are reproduced in multiple devices. Figure 2b is the data from device D2.

We investigate the thermal melting behavior of the stripe state at $\nu$ = 1/2 in Fig. 3. We evaluate the order parameter, or equivalently, $\theta$ (Fig. 3a upper panel) by anti-symmetrizing the $RC$ signal of Fig. 2a, $[RC(\phi) - RC(-\phi)] = 4\theta/\phi$, to remove the $\theta^2$-term. The anisotropy shows a strong temperature dependence and disappears around 35 K. This is in good agreement with the temperature scale at which the penetration capacitance peak at $\nu$ = 1/2 vanishes. To determine $\theta$ at $\nu$ = 1/2 more accurately, we use the optical response at the isotropic 2/3 state as a reference, that is, $R(\theta = 0)$, and measure $[RC(\phi) - RC(-\phi)]$ as a function of $1/\phi$ (Fig. 3b). A linear dependence is observed for all temperatures. We extract $\theta$ from the slope (Fig. 3c). Its uncertainty is dominated by sample drift in a helium cryostat and the spatially varying stripe domains (Fig. 4). The anisotropy decreases approximately linearly with temperature (Fig. 3c). This suggests that the stripe order is continuously melted by thermal fluctuations and it is potentially associated with a secondary order parameter[29].

We study the stripe domains at $\nu$ = 1/2 over the entire device D1 at 5 K in Fig. 4. To determine the anisotropy amplitude and orientation at each location, we measure the polarization rotation as a function of sample orientation and compare it to the theoretical dependence of $\theta = \theta_0 \sin[2(\alpha - \alpha_0)]$ (see supplementary notes 3 for derivation). Here $\theta_0$ is the anisotropy amplitude, and $\alpha$ and $\alpha_0$ denote the orientation of the sample and the anisotropy fast axis, respectively. We obtain a spatial map of $\theta$ for each sample orientation by following the design in Ref. [30] to achieve both high polarization purity and spatial resolution in a wide-field microscope mode.

Figure 4a shows an example at $\alpha$ = 16°. An anisotropy signal is observed in most of the regions of the device (inside the black dashed line), and its amplitude and sign vary spatially. If we rotate the sample by 90° (Fig. 4b), the sign of the signal reverses because the fast and slow axes are interchanged. Figure 4c illustrates the excellent fitting quality for three representative points on the device, labeled P1 to P3 in Fig. 4a. A map of the fitting parameter $\theta_0$ and $\alpha_0$ is shown in Fig. 4e. The color map represents the amplitude. The length and orientation of the line segments at each position represent the local amplitude and orientation of anisotropy. Multiple domains of different stripe orientations are observed.

We analyze the frequency distribution of stripe orientation $\alpha_0$ over the entire device to obtain a histogram in Fig. 4d. The anisotropy (fast) axis varies over a large range of angle. The distribution shows, however, two almost-equally-preferred angles that differ by 90°. These angles match well the armchair and zigzag directions, respectively, of the superlattice (orange dashed lines in Fig. 4d), which are independently determined from the second-harmonic-generation measurement (supplementary notes 2)[17,18,24]. We propose in Fig. 4f two possible charge orders for the stripe state at $\nu$ = 1/2. The electrons form linear or zigzag stripes separated by charge-deficit regions in between, which, with three-fold rotational symmetry, give rise to six possible stripe orientations. Most of these stripe configurations are observed with varying domain sizes and frequencies. The domain patterns change considerably after each thermal cycle, while the anisotropy axes remain to align preferentially with the same two high-symmetry directions of the superlattice (see supplementary notes 8). This is likely caused by the presence of a small uniaxial strain in the device that helps to pin the stripe fluctuations for diffraction-limited optical detection without greatly affecting the intrinsic correlations. Introducing controlled strain[31] in future studies could help to shed more light on the formation of the stripe states and to reveal the coupling between the electronic and lattice degrees of freedom.

We now turn to the experimental results at filling factors away from $\nu$ = 1/2.

Interestingly, Fig. 2 and 3a show prominent asymmetry of the electronic anisotropy in $\nu$ about $\nu = 1/2$. The anisotropy is substantially stronger for $\nu < 1/2$ than $\nu > 1/2$. In the absence of quantum fluctuations or electron hopping, the behavior of the $\nu < 1/2$ and $\nu > 1/2$ regions is expected to be symmetric because they are equivalent by interchanging electrons and holes[32]. The observed asymmetry reflects the importance of quantum fluctuations on stripe order.

The anisotropy peak in doping density is also substantially wider than the PL or capacitance peaks (Fig. 2 and 3a). A careful examination of Fig. 2 shows that the enhancement in both the PL and electronic anisotropy is observed at several other filling factors $\nu = 1/4$ (purple dashed line), 2/5 (green) and 3/5 (orange). These are likely additional stripe crystal states at the assigned commensurate fillings (see supplementary notes 9 for possible charge configurations). Figure 3a also shows that the anisotropy survives partially into the compressible regions of the system, where the penetration capacitance falls to the background level. (Due to the high contact resistance at low temperature, reliable capacitance data below 20 K is not available.) These finite-temperature striped compressible states indicate the existence of electronic liquid crystal (e.g. nematic and smectic) phases at incommensurate fillings.

The stripe crystal and electronic liquid crystal states can be understood from a single-band extended Hubbard model on a triangular lattice in the flat band limit (see Methods). A minimum model with up to third-neighbor interactions produces incompressible charge-ordered ground states at a sequence of commensurate fillings as found in our measurements. In particular, the ground state at $\nu = 1/2$ is an electron crystal composed of linear or zigzag charge stripes (Fig. 4f) for $V_3/V_2 < 1/2$ and $V_3/V_2 > 1/2$, respectively, where $V_i$ is the i-th nearest-neighbor interaction. Our experimental results suggest that the $WSe_2/WS_2$ superlattice sits on the border of the two cases. The deviation from bare Coulomb interactions ($V_3/V_2 \sim 0.87$) is potentially due to dielectric screening and/or strain effects. The model also predicts incompressible stripe crystal states at $\nu = 2/5$ and 3/5. The 2/5 state can be viewed as the closest packing of domain walls between the $\nu = 1/3$ and $\nu = 1/2$ domains. Extending this concept to generic fillings, the ground state at $1/3 < \nu < 2/3$ is an incommensurate array of domain walls that are parallel to each other to avoid intersection (which costs additional energy) and is generally compressible. This picture largely explains our observation of rotational symmetry breaking over a large range of filling between $\nu = 1/3$ and $\nu = 2/3$.

The model only considers the ground state of a classical system and neglects thermal or quantum fluctuations. The prevalence of these fluctuations in the present system, as we have found, can drastically modify the stripe orders. Future low-temperature measurements with high compressibility sensitivity are needed to fully understand the microscopic origin of the observed electronic liquid crystal states. One possible scenario that has long been discussed in the context of high-temperature superconductors is the quantum melting of stripe crystals through the introduction of quantum fluctuations[1,33]. This can be investigated in 2D moiré superlattices with continuous electrostatic gating, as we have shown here. It would also be intriguing to investigate the interplay of charge stripe fluctuations with spin fluctuations, which can be conveniently probed owing to the strong spin-orbit coupling and the spin-valley selection rules in TMDs[17,18]. Our results thus demonstrate that semiconducting TMD moiré superlattices are a promising platform to study the cooperation and competition between the different electronic phases of matter.

**Methods**:

Device fabrication: The angle-aligned $WS_2/WSe_2$ devices were made from exfoliated van der Waals materials using a layer-by-layer dry transfer method with a polycarbonate (PC) stamp[34]. All atomically thin flakes, including $WSe_2$ and $WS_2$ monolayers, few-layer graphite, $TaSe_2$ and hBN, were exfoliated from bulk crystals onto Si substrates with a 300-nm oxide layer. The crystal orientations of monolayer $WSe_2$ and $WS_2$ flakes were determined by angle-resolved optical second-harmonic generation[17,18,24]. The flakes were aligned before the transfer according to the crystal orientations to result in either a 0°- or 60°-twist-angle stack. The two cases can be distinguished from the second-harmonic signal from the heterostructure after assembly[17,18,24]. The hBN flakes of about 20-30 nm in thickness were used as gate dielectric layers. Few-layer graphite or $TaSe_2$ flakes were used as gate electrodes. Electrical contacts to the $WS_2/WSe_2$ bilayers were achieved by using few-layer $TaSe_2$. The assembled stacks were released at about 180°C onto Si substrates with pre-patterned Ti/Au electrodes. The residual PC on the device surface was dissolved in chloroform, followed by rinsing the device in acetone and isopropyl alcohol. We have measured multiple devices of both 0° and 60° twist angles. Optical anisotropies from stripe phases were observed only in 0°-twist-angle devices.

Electrical capacitance measurement: The penetration capacitance of the $WSe_2/WS_2$ heterostructures as a function of doping density was measured in a close-cycle He-4 cryostat (Oxford TeslatronPT). It is an alternative to electrical transport measurements, which are difficult in TMD moiré superlattices at low temperature because of the high contact resistance. A reasonably good electrical contact to the $WSe_2/WS_2$ bilayers was achieved by applying a large positive top gate voltage to heavily dope the contact region. A back gate voltage was swept to tune the electron density in the $WSe_2/WS_2$ bilayers. The back gate was designed to avoid the contact region (Fig. 1b) so that it does not affect the contact resistance significantly throughout the measurement. To measure the relatively small change in the device capacitance, a commercial high electron mobility transistor (HEMT, model FHX35X) was used to effectively disconnect the small device capacitance from the large parasitic capacitance[35]. The HEMT was connected to the top gate; the ac voltage drop across its source and drain terminals was measured by a lock-in amplifier while an ac excitation voltage was applied to the back gate. Measurements were performed at 6.923 kHz with a 10 mV excitation voltage. A circuit diagram is provided in supplementary notes 1. In the low frequency and low resistance limit, the penetration capacitance is given by $C_P = C_{TG}C_{BG}/(C_{TG} + C_{BG} + C_q)$, where $C_{TG}$ ($\approx C_{BG} = C_G$) is the geometric capacitance between the $WSe_2/WS_2$ bilayer and the top (bottom) gate electrode, and $C_q$ is the sample quantum capacitance. A small (large) value of $C_P$ indicates compressible (incompressible) state. The penetration capacitance data shown in the main text are normalized by the gate capacitance.

Calibration of the doping density and filling factor: The filling factor $\nu$ is defined as the number of electrons per moiré superlattice unit cell. The lattice constant of angle-aligned $WSe_2/WS_2$ moiré superlattices is about 8 nm, which corresponds to an electron density of $1.9 \times 10^{12}$ $cm^{-2}$ for $\nu = 1$. The conduction band edge was determined from the sharp drop of the penetration capacitance signal (Fig. 1d). The electron density was estimated from the gate voltage based on the parallel-plate capacitor model with known hBN thickness (determined by AFM measurements) and dielectric constant (~ 3, Ref. [18]). We first assigned the $\nu = 1$ and $\nu = 2$ insulating states, which were used as landmarks to obtain the conversion factor between the gate voltage and $\nu$. With this conversion factor we assigned other features in both the capacitance and PL measurements to the closest rational number with a small denominator, such as 1/3, 1/2, 2/3, etc. The PL measurement further allows us to determine the local filling factor at each position and investigate the spatial inhomogeneity in carrier density. We estimate $\Delta\nu/\nu$ <10%, which is also demonstrated by the narrow peak widths of the incompressible states in PL and penetration capacitance measurements (Fig. 2 and 3a).

<u>Optical anisotropy measurement</u>: A single-wavelength continuous-wave probe light from a Ti-Sapphire laser (M Squared SOLSTIS system) was used to measure the optical birefringence of the devices (Fig. 1e). The photon energy was set to 1.685 eV (736 nm) and 1.676 eV (740 nm), respectively, for device D1 and D2 to match their optical resonances for enhanced detection sensitivity. The light intensity was 5 nW/$\mu m^2$ to minimize the heating effects. A broadband half-wave plate was set before analyzer P2 to effectively change its angle $\phi$. An 800-nm short-pass filter was set after P2 to remove the heterostructure's PL. The reflected light was detected by a 2D electron-multiplying charge-couple device (CCD) camera (Princeton Instruments, ProEM 512x512). In the wide-field imaging mode, a lens was set before the beam splitter to focus the incident light on the back aperture of the objective. This ensures a small input numerical aperture and high polarization purity, as has been detailed in Ref. [30]. The spatial resolution of these measurements is about 800 nm.

<u>PL measurement</u>: Identical experimental configurations were used for the PL and the optical anisotropy measurement in Fig. 2 (see above). For the PL measurement, the 800-nm short-pass filter was replaced by an 800-nm long-pass filter. Figure 2 shows the result at 12.5 K to maximize the visibility of the enhancement at filling 1/4 and 2/5. The PL spectrum in supplementary notes 4 was measured using a different configuration. In this case, the sample was illuminated by a 532-nm excitation and the PL emission was guided into a monochromator and detected by a liquid-nitrogen-cooled CCD camera.

<u>Sample orientation dependence</u>: To obtain the stripe domain orientations, we measured $\theta$ as a function of sample orientation. A Soleil-Babinet compensator (Thorlabs SBC-VIS) was added right before the sample. The compensator was tuned and aligned to provide precise half-wave retardance at the probe wavelength, i.e. to function as a half-wave plate. We rotated the compensator to vary the angle $\alpha$ between the incident light polarization and the sample in the *x*-*y* plane. The reflected light goes through the compensator second time, and the light polarization is reverted to the original one. Rotating the compensator is equivalent to rotating the sample because the light polarization is changed only on the sample but not anywhere else.

Extended Hubbard model: To intuitively understand the observed stripe phases, we consider an extended Hubbard model on a triangular lattice:

$$H = -t\sum_{\langle ij\rangle}\left(c_i^{+}c_j + h.c.\right) + U\sum_i n_{i\uparrow}n_{i\downarrow} + \frac{1}{2}\sum_{i\neq j} V_{ij}n_i n_j. \quad (1)$$

The three terms correspond to hopping, on-site Coulomb interaction and long-range inter-site Coulomb interactions, respectively. In the following analysis, we consider the classical limit $t \to 0$ and the large $U$ limit $U \to \infty$. Under such an approximation, states at filling factor $\nu < 1$ are restricted from having more than one charge per site, that is, each site is either empty ($n_i = 0$) or singly occupied ($n_i = 1$). The ground state is therefore solely determined by the long-range interactions $V_{ij}$.

We start from the first three terms in $V_{ij}$, $V_1$, $V_2$, $V_3$, which correspond to the first, second, and third nearest-neighbor repulsions. This model has been analyzed in Ref. [36] at several commensurate fillings. The ground state at $\nu = 1/3$ is an isotropic electron crystal, which is consistent with our observation. The ground state at $\nu = 1/2$ features linear or zigzag stripes depending on the ratio of $V_3/V_2$, both explicitly breaking the three-fold rotational symmetry. The model also predicts that the $\nu = 1/4$, 2/5 and 3/5 states break rotational symmetry. Interestingly, the charge configurations at $\nu = 2/5$ can be obtained by tightly packing domains of the $\nu = 1/3$ state and $\nu = 1/2$ states.

Following this picture, states at a generic filling $1/3 < \nu < 1/2$ (and similarly $1/2 < \nu < 2/3$) can be obtained by adding domain walls – instead of individual particles – to the electron crystal at $\nu = 1/3$ and 1/2. This allows us to include longer-range interactions perturbatively and estimate the compressibility of the system. At a generic filling, the ground state is an incommensurate array of domain walls[37]. The spacing between walls is determined by minimizing the total energy including the domain-wall energy and the interaction energy between walls. At doping $\nu = 1/3 + \delta$ or $1/2 - \delta$ ($\delta > 0$), where the walls are far apart, the ground-state energy per unit length perpendicular to the wall takes the form[38]:

$$\frac{E}{l} = -\left(\frac{q}{l}\right)\delta\mu + \left(\frac{\lambda}{l}\right)e^{-\kappa l}. \quad (2)$$

Here $l$ is the average separation of domain walls, $q$ is the net charge per unit length along the wall, $\mu$ is the chemical potential, and $\lambda$ and $\kappa$ describe the interaction between walls. This interaction falls off exponentially over distance, with $\kappa$ setting the range of the screened Coulomb repulsion. Minimizing the energy yields $l \sim \log(1/\delta\mu)$, which varies continuously with the chemical potential. The compressibility is then given by

$$\frac{\partial n}{\partial \mu} \sim \frac{1}{\log(1/\delta\mu)}, \quad (3)$$

which vanishes non-analytically at the transition from the commensurate crystal to the incommensurate domain-wall array. In this picture, the rotational symmetry breaking exists throughout the filling range $1/3 < \nu < 2/3$, which provides a potential explanation of our observation.

On the other hand, the above model only applies to the ground state in the classical limit while both quantum and thermal fluctuations play an important role in our system. For example, the model predicts symmetric behaviors between the $\nu < 1/2$ and $\nu > 1/2$ regions, while a prominent asymmetry is observed experimentally. The quantum effects can drastically modify the stripe orders, such as drive the stripes to fluctuate at zero-temperature and quantum-melt a stripe crystal into a quantum liquid crystal[1,33]. These intriguing scenarios can be explored by investigating the characteristic filling-dependent compressibility at low temperature and with high sensitivity.

**Data availability**:

The data that support the findings of this study are available within the paper and its Supplementary Information. Additional data are available from the corresponding authors upon request.

**References for Methods**:

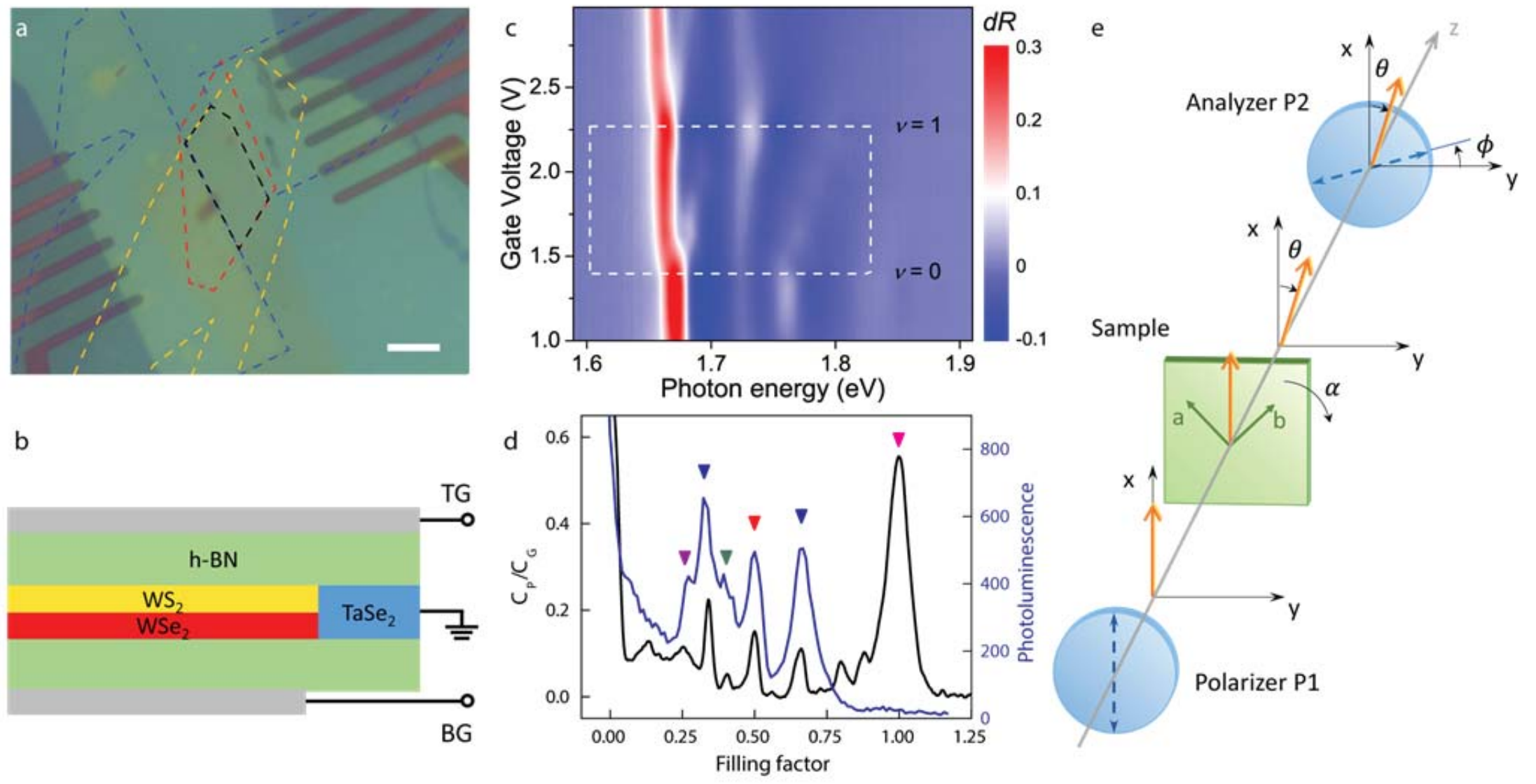

**Figure 1 │ Optical detection of electronic anisotropy. a, b**, Optical microscope image (**a**) and side-view illustration (**b**) of near-zero-twist-angle $WSe_2/WS_2$ heterostructure D1. Blue, red, yellow and black dashed lines in (**a**) are contours of $TaSe_2$, $WSe_2$, $WS_2$ and the $WSe_2/WS_2$ flakes, respectively. Scale bar is 5 μm. Gate voltages are applied on both the top (TG) and bottom (BG) gate to tune the carrier density. **c**, Gate-dependent differential reflectance spectrum of the heterostructure. Dashed box marks the electron doping region between $\nu = 0$ and 1. **d**, Penetration capacitance at 20 K (black curve) reveals incompressible states at $\nu = 1$ (magenta arrow), 1/3 and 2/3 (blue arrows), 1/2 (red arrow), 1/4 (purple arrow) and 2/5 (green arrow). The interlayer exciton PL intensity (blue curve) shows enhancement at these insulating states in the low doping regime. $C_P$ and $C_G$ are penetration capacitance and geometric capacitance, respectively (see methods). **e**, Schematics of optical anisotropy measurement. An anisotropic sample with fast and slow axis $a$ and $b$, respectively, is placed between polarizer P1 (transmission along the $x$-axis) and analyzer P2 (transmission at angle $\phi$ from the $y$-axis). Sample-induced rotation $\theta$ of the incident light polarization is detected from the intensity change after P2. The schematics show a transmission geometry for simplicity while a reflection geometry is used in real experiment.

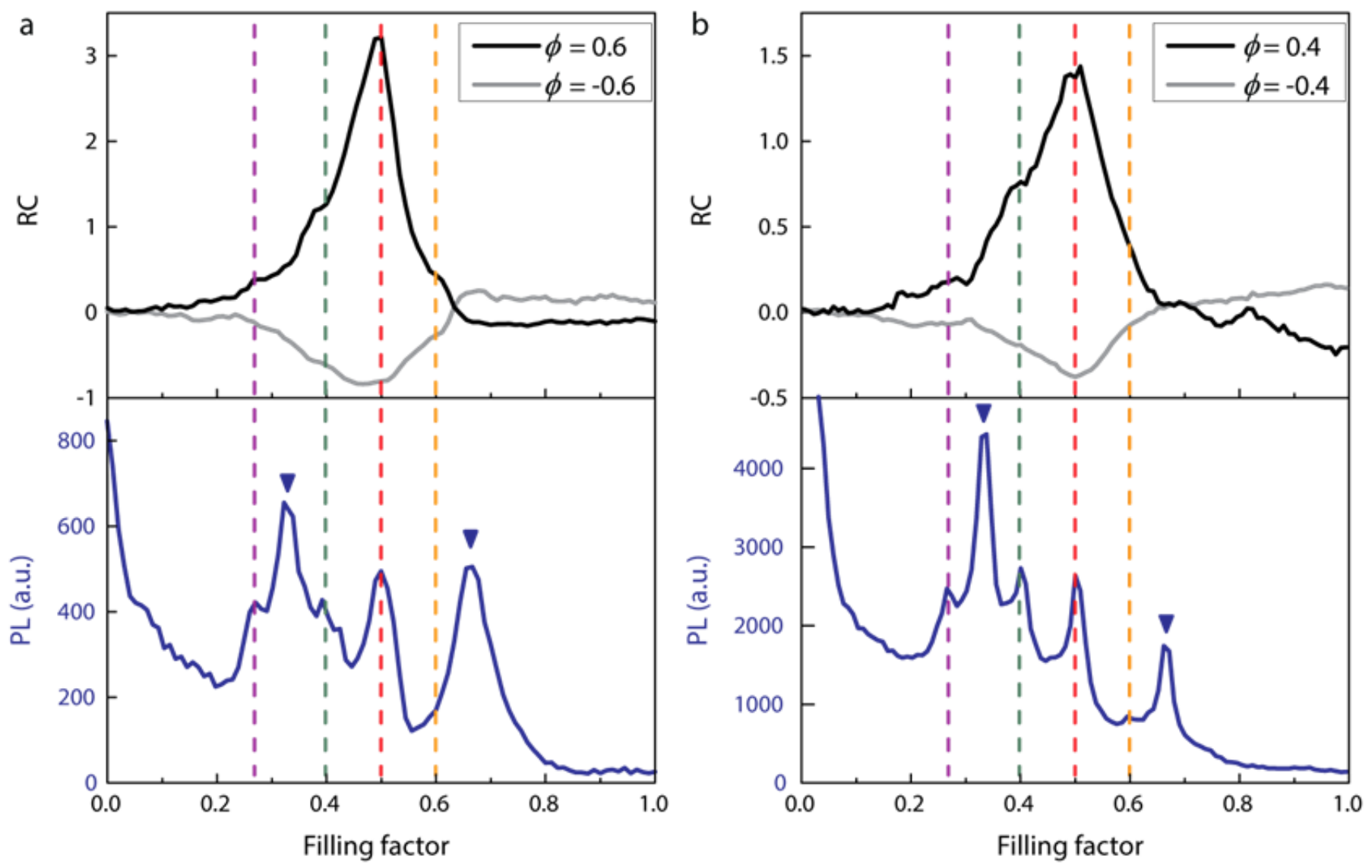

**Figure 2 | Electronic anisotropy in $WSe_2/WS_2$ moiré superlattices. a**, (Top) Doping dependence of the relative intensity change *RC* (defined in the text) of device D1 at 5 K. The probe is at 1.685 eV. The black and grey lines are measurements at $\phi$ = 0.6° and -0.6°, respectively. (Bottom) The corresponding interlayer exciton PL intensity excited by the probe light. The electronic anisotropy peaks at $\nu = 1/2$ (red dashed line), which coincides with a PL intensity peak. It is assigned to a stripe crystal state. Additional stripe crystal states are identified at commensurate fillings of 1/4 (purple), 2/5 (green) and 3/5 (orange), at which enhancement in both the electronic anisotropy and PL intensity is observed. The anisotropy signal shows marked asymmetry around $\nu = 1/2$, suggesting the importance of quantum fluctuations. **b**, Same as (**a**) for device D2. The probe is at 1.676 eV. Similar behaviors are observed.

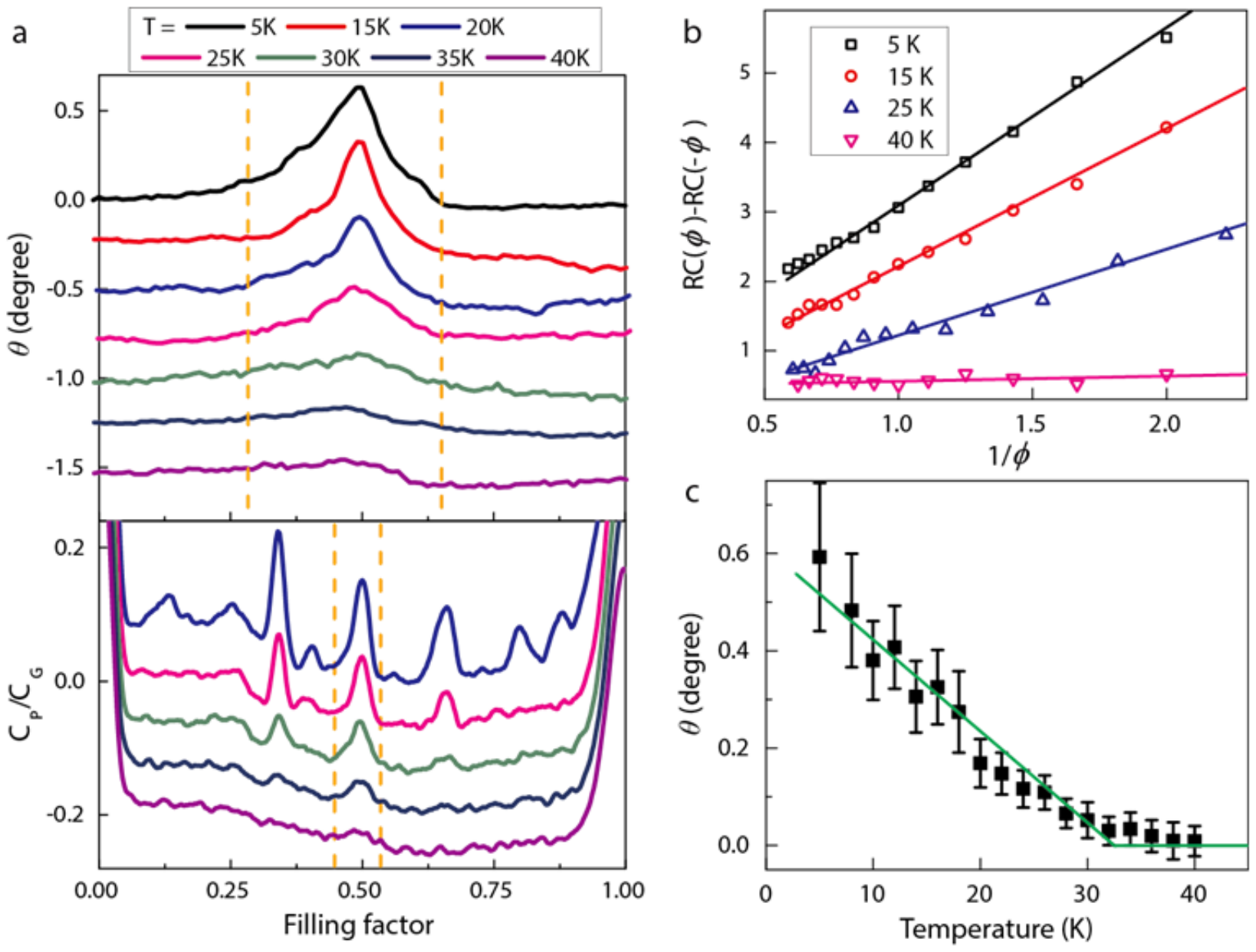

**Figure 3 | Temperature dependence. a**, Gate-dependent polarization rotation at different temperatures (upper panel) for device D1 and the corresponding penetration capacitance (lower panel). In both measurements the peak at $\nu$ = 1/2 decreases continuously with temperature and disappears around 35 K, indicating continuous thermal melting of the stripe crystal. Vertical dashed lines mark the region with optical anisotropies (upper panel) and the upper limit of the insulating region around $\nu$ = 1/2 (lower panel). The optical anisotropies persist into the metallic region. The curves for different temperatures are vertically displaced for clarity. **b**, Dependence of the anisotropy signal as a function of analyzer angle $\phi$ at $\nu$ = 1/2 for several representative temperatures (symbols). The polarization rotation angle $\theta$ is obtained from the slope of the linear fittings (lines). **c,** Temperature dependence of $\theta$ at $\nu$ = 1/2. Green solid line shows a linear fitting at low temperature and a zero-rotation baseline at high temperature. Error bars are the standard deviation of $\theta$'s from 3×3 pixels (~ 1×1 μm), which is the typical scale of sample drift in the temperature dependence measurement.

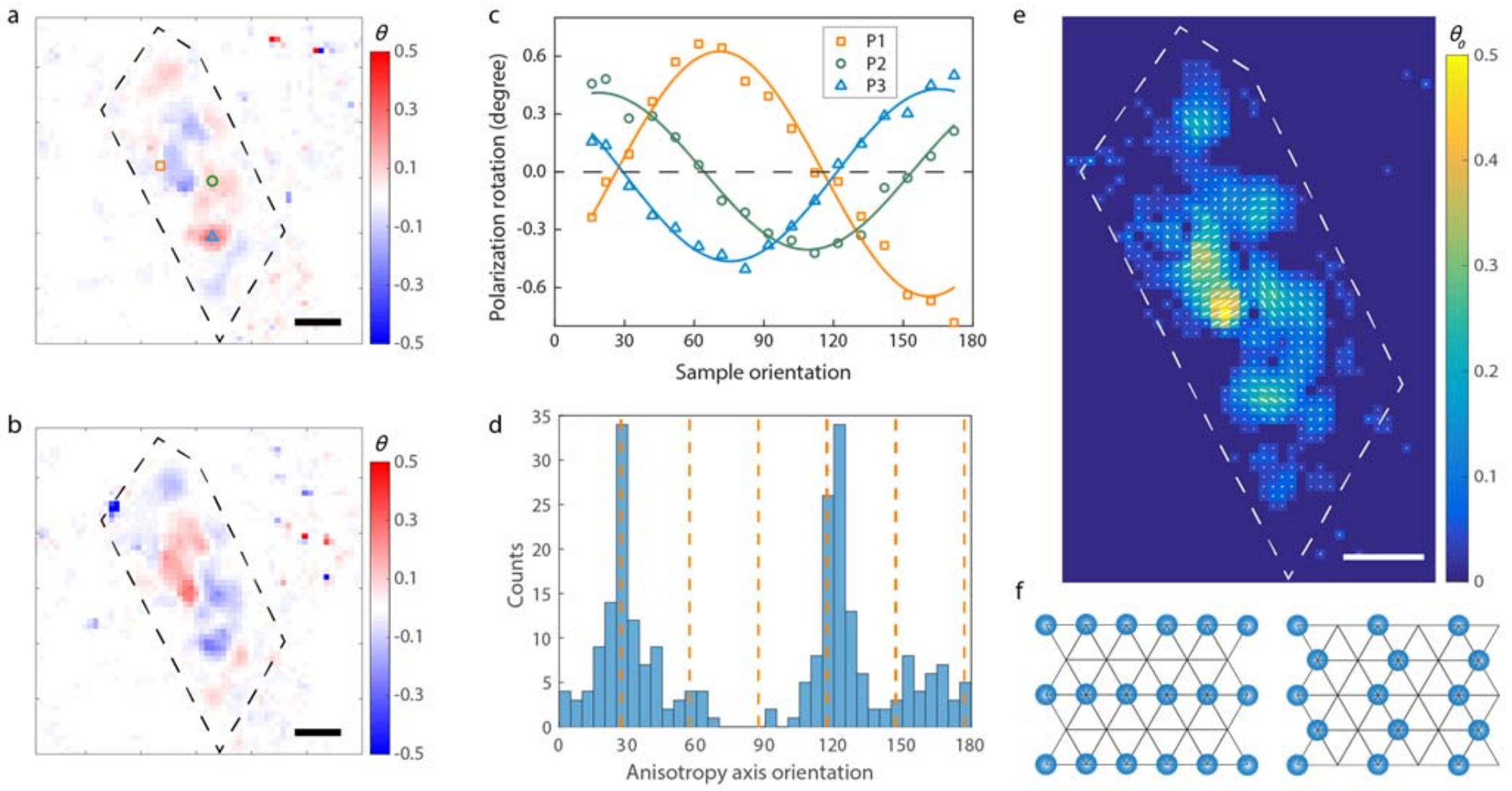

**Figure 4 | Stripe domains at $\nu$ = 1/2.** **a**, **b**, Spatial maps of $\theta$ (in degrees) at two sample orientations that differ by close to 90°, $\alpha$ = 16° (**a**) and 102° (**b**). Dashed line shows the contour of device D1. **c,** Sample orientation-dependent $\theta$ at three representative points (labeled P1 to P3 in (**a**)). Symbols are experimental data; lines are fits to the theoretical relation of $\theta = \theta_0 \sin[2(\alpha - \alpha_0)]$. Horizontal dashed line marks zero rotation. **d**, Distribution of fast axis direction over device D1 shows two preferred angles that match well the high-symmetry directions of the superlattice (vertical dashed lines). **e**, Domain pattern of electronic stripes. The length and orientation of the line segments at each point represent the local amplitude and orientation of anisotropy; the color corresponds to the anisotropy amplitude. **f**, Two possible charge orders with a linear and zigzag stripe for the $\nu$ = 1/2 state that are consistent with the result of (**e**). Scale bars in **a**, **b**, and **e** are 2 μm.

**Supplementary Information for**

**Stripe phases in $WSe_2/WS_2$ moiré superlattices**

Chenhao Jin†*, Zui Tao†, Tingxin Li†, Yang Xu, Yanhao Tang, Jiacheng Zhu, Song Liu, Kenji Watanabe, Takashi Taniguchi, James C. Hone, Liang Fu*, Jie Shan*, Kin Fai Mak*

† These authors contributed equally to this work

* Correspondence to: jinchenhao@cornell.edu, liangfu@mit.edu, jie.shan@cornell.edu, km627@cornell.edu

1. **Electrical capacitance measurement**
2. **Determination of crystal axes**
3. **Relation between electronic anisotropy and polarization rotation**
4. **Photoluminescence spectra of device D1**
5. **Temperature-depend differential reflectance of device D1**
6. **Reproducibility of the gate-dependent anisotropies**
7. **Control experiments (probe intensity and photon energy dependence)**
8. **Thermal cycle effects on stripe domain patterns**
9. **Possible charge configurations at other commensurate fillings**

## 1. Electrical capacitance measurement

Figure S1 illustrates the circuit diagram of the penetration capacitance measurements, which has been detailed in the Methods section of the main text. Here $V_T$, $V_B$ and $V_{ac}$ are the top gate, bottom gate and ac excitation voltage, respectively. In addition, $V_H$ and $I_H$ are applied to set the work point of the high electron mobility transistor (HEMT).

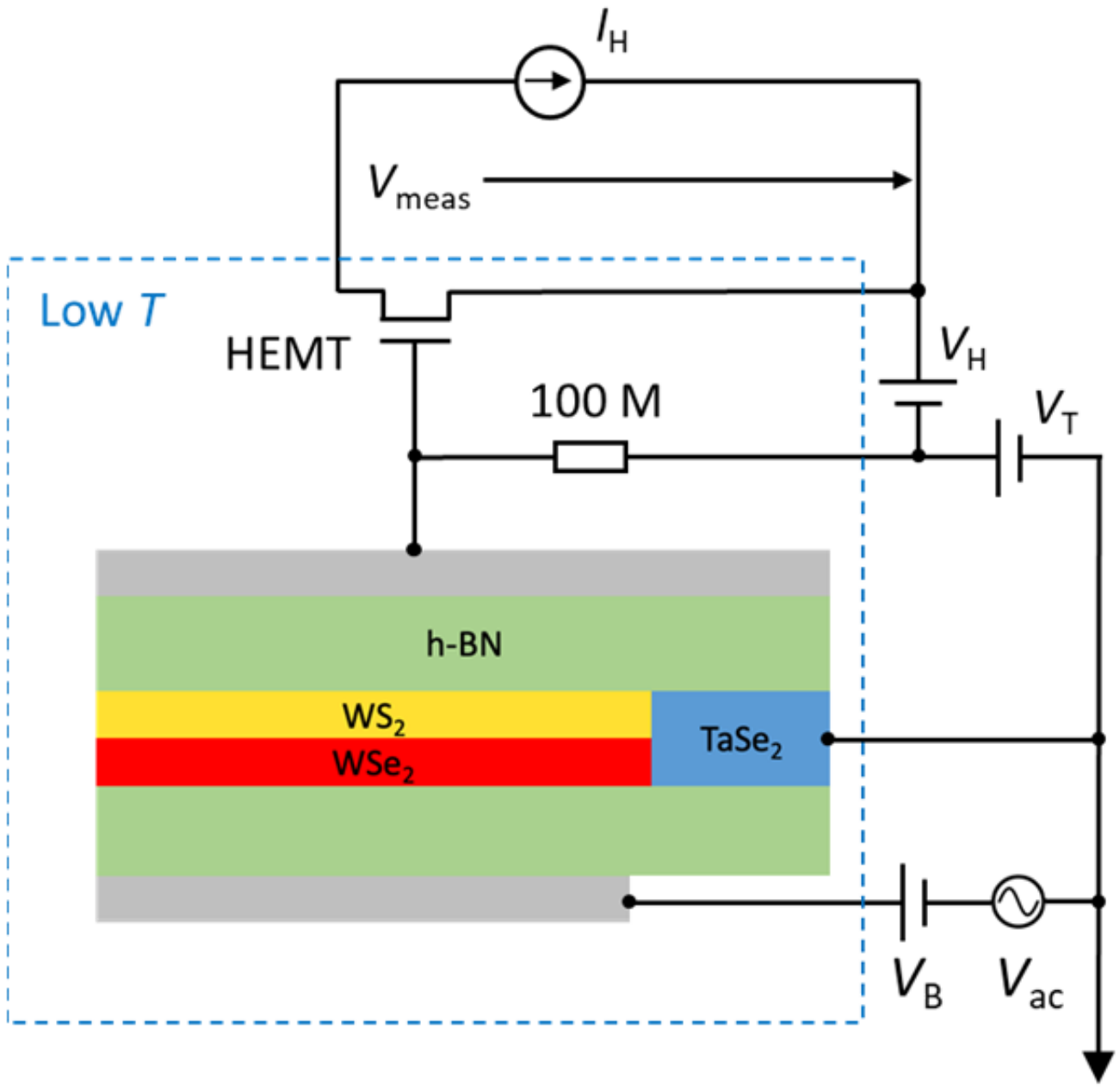

Figure S1. Schematic drawing of the penetration capacitance measurement. Dashed box encloses components inside the cryostat.

## 2. Determination of crystal axes

Figure S2 shows the second harmonic generation (SHG) signal from the $WSe_2$ and $WS_2$ layers in device D1 as a function of sample orientation $\alpha$. The polarization of the fundamental radiation (800 nm) is perpendicular to the collection polarization of the second-harmonic radiation (400 nm). The peaks correspond to the armchair directions of the crystal.

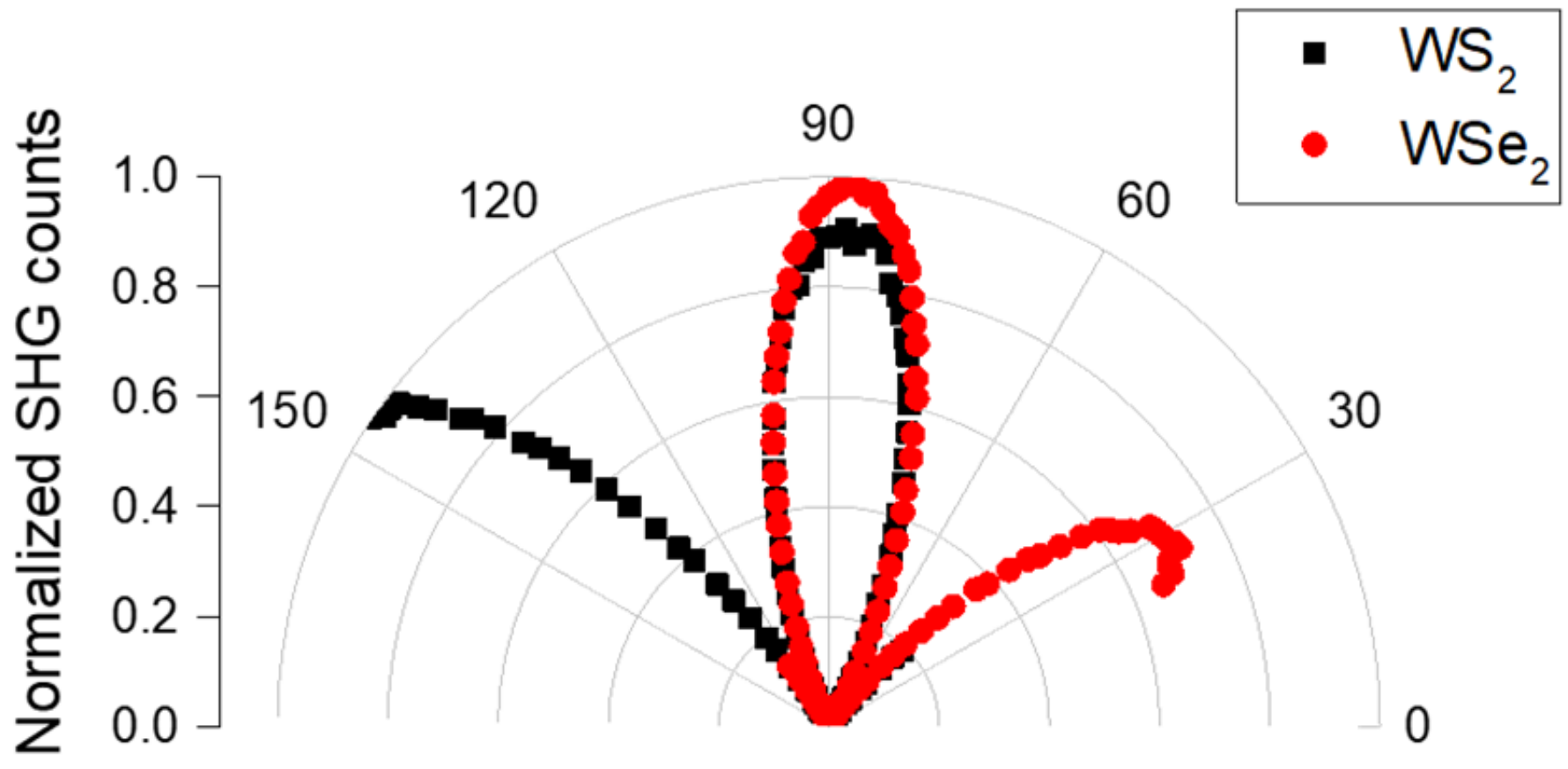

Figure S2. Second harmonic generation from device D1 as a function of sample orientation $\alpha$.

### 3. Relation between electronic anisotropy and polarization rotation

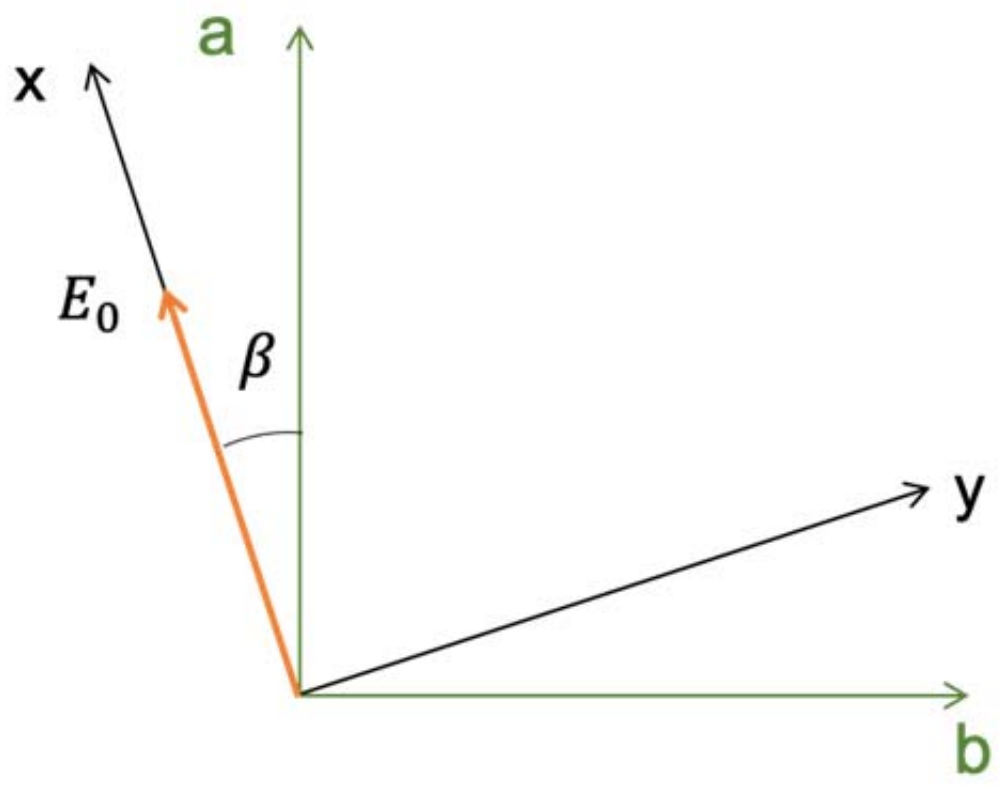

Figure S3. A linearly-polarized light is reflected from an anisotropic sample with anisotropy axis $a$ and $b$. The incident light polarization (orange arrow) is along the *x*-axis that has an angle $\beta$ relative to the $a$ axis.

In an anisotropic two-dimensional (2D) sample, the optical conductivity tensor can be written as $\sigma = \mathrm{diag}\{\sigma_a, \sigma_b\}$, where $\sigma_a$ and $\sigma_b$ are the optical conductivity along the two anisotropy axes $a$ and $b$ (Fig. S3). The electronic anisotropy can be characterized by an order parameter $\eta$, which describes the difference between $\sigma_a$ and $\sigma_b$ through $\sigma_a = \sigma_b + \eta$. Here $\sigma_a, \sigma_b$ and $\eta$ are generally complex and we will assume $\eta$ to be small in the following discussion.

Consider light reflection from the sample ($WSe_2/WS_2$ bilayer) on an isotropic substrate (hBN-$SiO_2$-Si). In the 2D limit, the sample's contribution to the light reflection can be treated as a perturbation. The reflection coefficients along the $a$ and $b$ axes can be expressed as[30]:

$$r_a = r_0(1 + L\sigma_a),\ r_b = r_0(1 + L\sigma_b)\,. \quad \text{(S1)}$$

Here $r_0$ is the reflection coefficient of the bare hBN-$SiO_2$-Si structure without the sample, and $L$ is a constant that depends on the local field factor from the hBN-$SiO_2$-Si structure.

When a linearly-polarized light impinges on the sample of amplitude $E_0$ and polarization angle $\beta$ (Fig. S3), the electric field of the reflected light along the *a* and *b*-axis is:

$$E'_a = r_a E_0 \cos\beta\,, E'_b = -r_b E_0 \sin\beta. \quad \text{(S2)}$$

The components of reflected light parallel and perpendicular to the incident light polarization are respectively:

$$E'_x = -E'_b \sin\beta + E'_a \cos\beta \approx E_0 r_a,$$
$$E'_y = E'_b \cos\beta + E'_a \sin\beta = E_0 \sin\beta \cos\beta (r_a - r_b). \quad \text{(S3)}$$

The difference in $r_a$ and $r_b$ results in a finite electric-field component perpendicular to the incident polarization direction, which consequently is a polarization rotation. We thereby obtain the polarization rotation $\theta$ as:

$$\theta = \frac{E_0 \sin\beta \cos\beta (|r_a| - |r_b|)}{E_0\ |r_b|}$$
$$= \frac{|r_0|^2}{|r_b|^2} Re[L(\,\sigma_a - \sigma_b)] \sin 2\beta \ = \frac{1}{|1 + L\sigma_b|^2} Re[L\eta] \sin 2(\alpha - \alpha_0). \quad \text{(S4)}$$

In the last step we have used the relation $\beta = \alpha - \alpha_0$, where $\alpha$ and $\alpha_0$ are the orientation of the sample and the anisotropy fast axis *a* measured in the *x-y* frame. For a fixed sample orientation $\alpha$, Eq. S4 indicates that the polarization rotation $\theta$ is proportional to the electronic anisotropy $\eta$. Therefore, $\theta$ can be directly used as the order parameter to study the doping and temperature dependence. On the other hand, when the sample is rotated, $\theta$ should show a characteristic dependence of $\sin 2(\alpha - \alpha_0)$. This allows us to determine the anisotropy axes *a* and *b*.

### 4. Photoluminescence spectra of device D1

Figure S4 shows the gate-dependent photoluminescence (PL) spectrum of device D1 at 15 K. The excitation light has a wavelength of 532 nm (2.33 eV) and an intensity of 10 nW/μm$^2$. In the absence of doping, the prominent PL peak at ~ 1.41 eV originates from the interlayer exciton emission of the $WSe_2$/$WS_2$ moiré superlattice[24]. In addition, the PL intensity shows clear enhancement at fillings 1/3, 1/2 and 2/3, consistent with the results described in the text when the device is excited at the lowest-energy moiré exciton resonance 1.685 eV (736 nm). This confirms that the observed PL enhancement at particular fillings is not due to gate-dependent optical absorption change, because the excitation light here has a much higher energy than any specific exciton resonances.

In addition, the PL spectra at $\nu$ = 1/3, 1/2 and 2/3 are rather similar to that at charge neutrality. This indicates that PL at these fillings has the same origin, that is, from the interlayer exciton emission. The PL enhancement can be understood from the fact that PL has the highest intensity at charge neutrality ($\nu$ = 0, see Fig. 2 in the text), and a gapped state such as $\nu$ = 1/3, 1/2 and 2/3 can partially restore the charge neutrality behavior. This scenario also explains the similarity between the PL spectra at charge neutrality and at the incompressible states at fractional fillings.

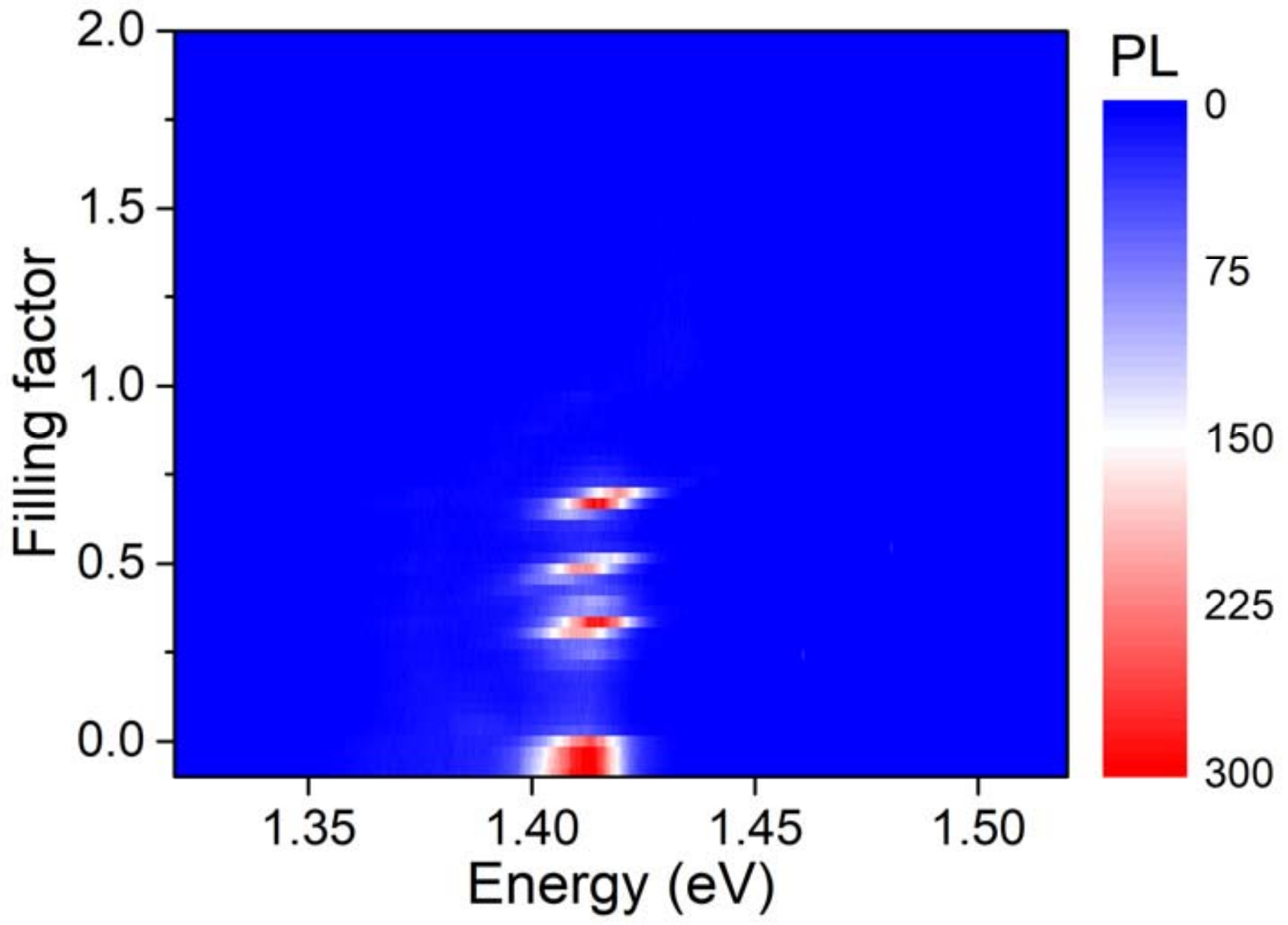

Figure S4. Filling-dependent PL spectrum of device D1 at 15 K.

**5. Temperature-dependent differential reflectance of device D1**

Figure S5 shows the differential reflectance of device D1 at temperatures of 5, 15, 30 and 40 Kelvin. The reflectance spectra remain largely unchanged except for a slight redshift of < 2 meV. Therefore, the temperature-dependent optical anisotropies in Fig. 3 of the main text cannot originate from the temperature-dependent optical resonance. Instead, it is caused by thermal melting of the stripe phases.

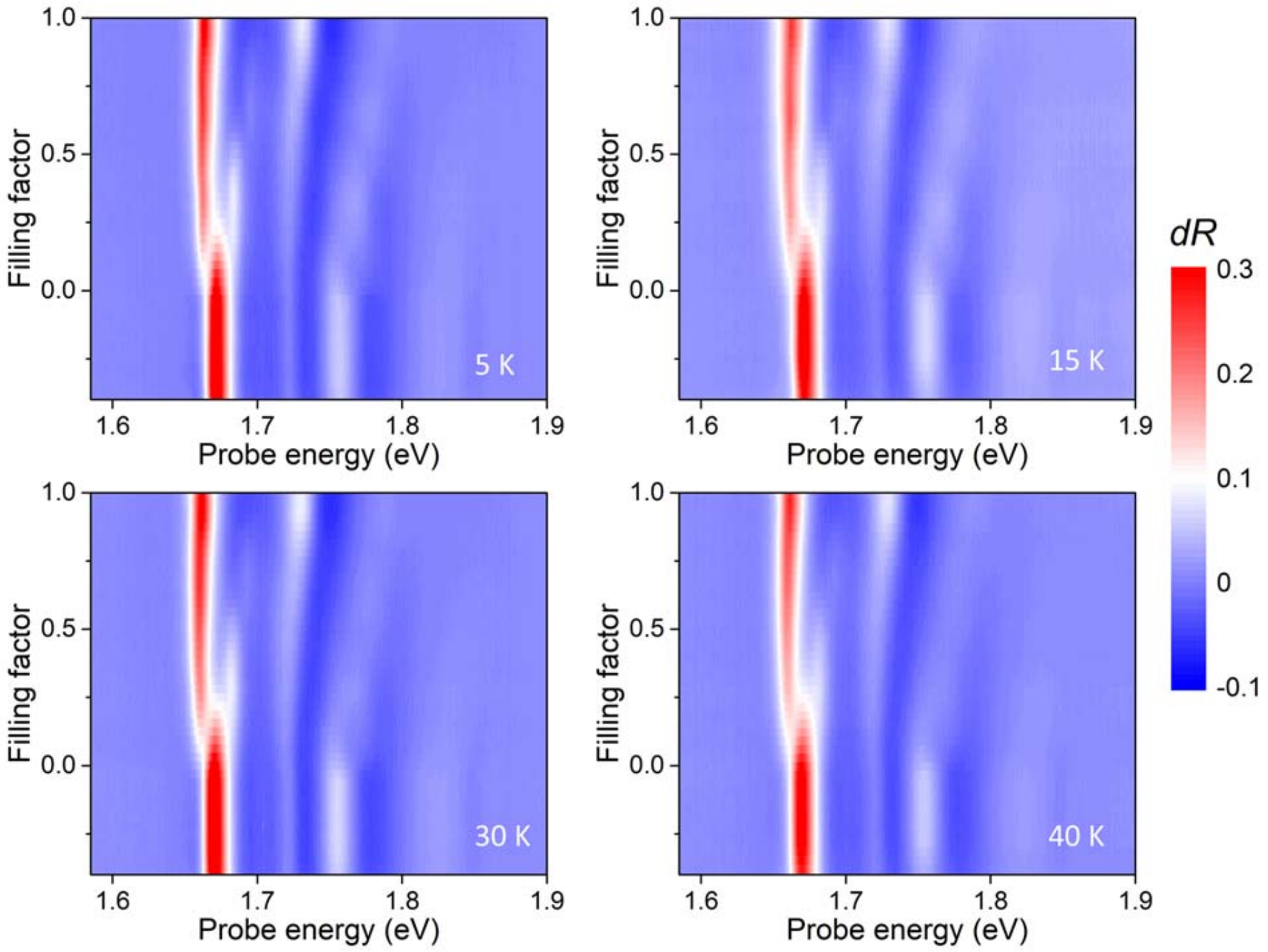

Figure S5. Temperature-dependent differential reflectance spectrum of device D1. The spectra remain largely unchanged over the whole temperature range (5 - 40 K) except for a slight redshift.

## 6. Reproducibility of the gate-dependent optical anisotropies

The observed anisotropy profile is well reproducible within a single device as well as across multiple devices. Figure 2 in the text has demonstrated the consistent behaviors between device D1 and D2. In addition, Figure S6 shows the doping-dependent polarization rotation $\theta$ of four representative positions on device D1. The maximum polarization rotation is normalized to 1. The sample orientation is fixed at $\alpha = 72°$. All points show a prominent electronic anisotropy peak at $\nu = 1/2$ (red dashed line) and enhancement at several additional fillings $\nu = 1/4$ (purple dashed line), 2/5 (green) and 3/5 (orange). In addition, the anisotropy profile in doping-dependence is well reproduced. Particularly, they all show marked asymmetry between $\nu < 1/2$ and $\nu > 1/2$, indicating the importance of quantum fluctuations.

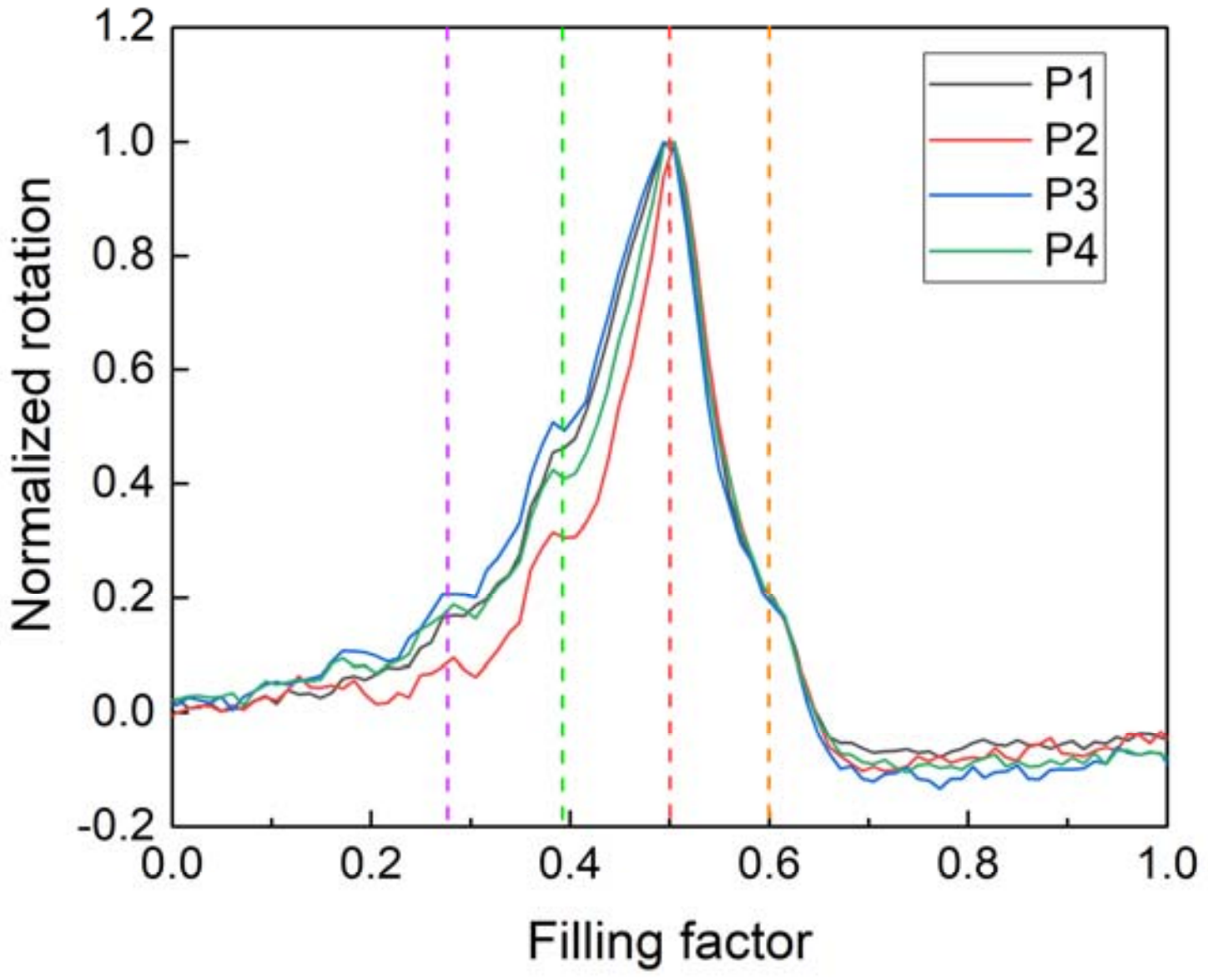

Figure S6. Doping-dependent polarization rotation $\theta$ of 4 different spatial points on device D1. The maximum absolute rotation is normalized to 1. Both the enhanced anisotropy at commensurate fillings and the anisotropy peak profile are well reproducible.

**7. Control experiments (probe intensity and photon energy dependence)**

We have performed several control experiments to verify that the observed optical anisotropies arise intrinsically from the electronic stripe phases:

(i) Probe light intensity dependence.

In our experiment, we probe the linear response of the correlated states in moiré superlattices. To this end, we have performed a systematic probe intensity dependence study. Figure S7 shows the polarization rotation $\theta$ on device D2 measured with three different probe light intensities 2.5, 5.0 and 10.0 nW/μm$^2$. No intensity dependence is observed in the anisotropy signal, confirming that we are in the linear regime and the probe light's perturbation to the system is negligible. Throughout the measurement we have kept the probe light intensity at 5 nW/μm$^2$.

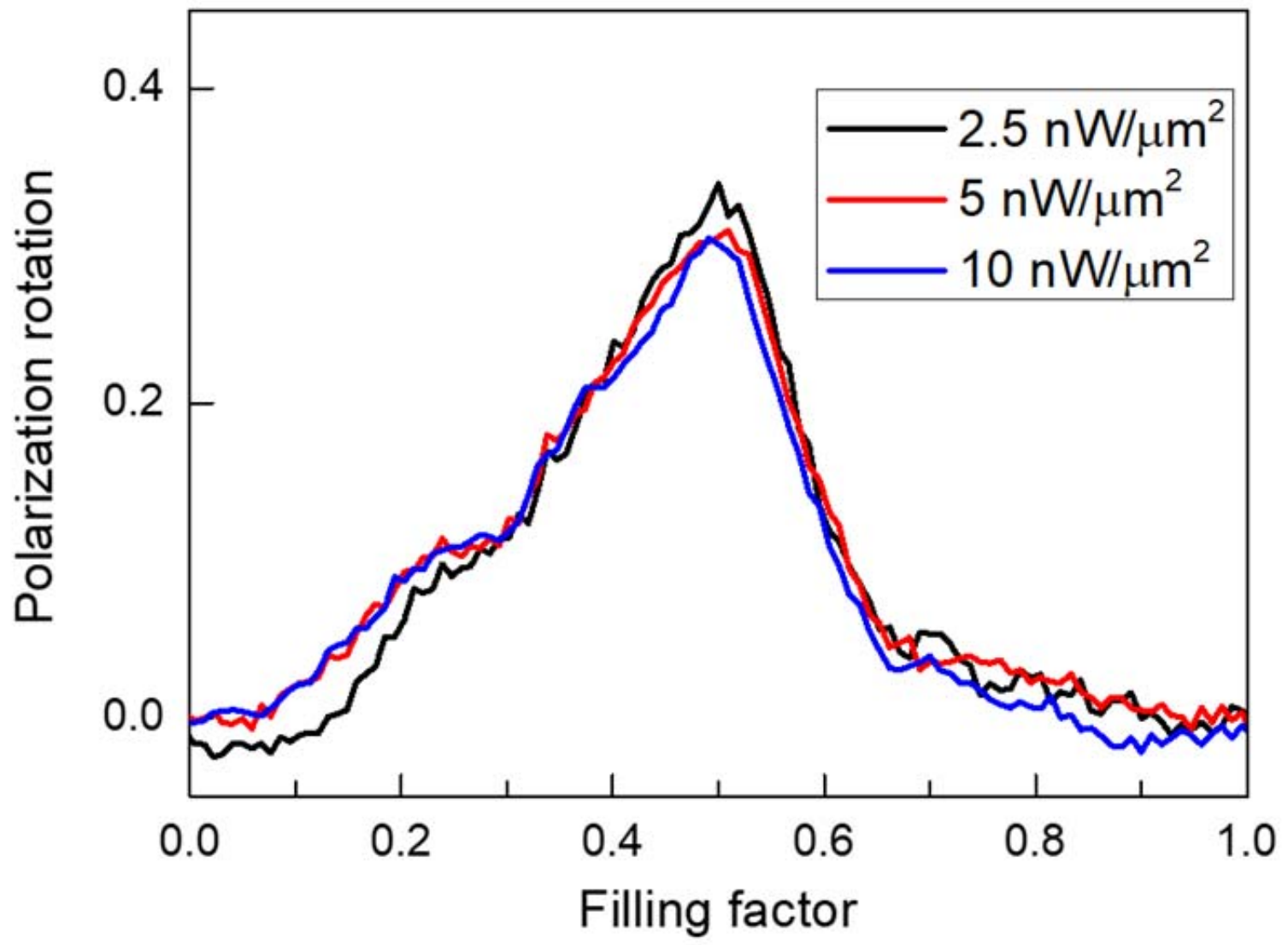

Figure S7. Probe light intensity dependence. The anisotropy signal does not change with probe light intensity, indicating that the probe is in the linear regime and the probe does not significantly perturb the correlated states.

(ii) Probe photon energy dependence

We consider potential contributions to the gate-dependent anisotropy signal from the gate-dependent optical resonances. The simple optical reflectivity change does not contribute to the anisotropy signal since the anisotropy is determined from the slope of the $\phi$ dependence of the reflected intensity (Fig. 2 and Fig. 3b). Here $\phi$ is the polarizer angle and the sample reflectivity does not depend on $\phi$. However, the relation between the electronic anisotropy and the measured probe polarization rotation depends on the probe photon energy (Eq. S4). Below we show that the profile of the gate-dependent anisotropy signal does not depend on the probe wavelength although the overall amplitude does.

Figure S8a shows the distinct gate dependences of the optical reflection from device D1 at three different photon energies 1.692, 1.685 and 1.671 eV (denoted by arrows in Fig. S8b). It is measured without analyzer P2. The optical reflection change is normalized to [0,1] for each probe photon energy. In contrast, the filling dependence of normalized $\theta$ (Fig. S8c) is nearly photon energy independent. Figure S8d shows the absolute value of $\theta$ at $\nu = 1/2$ as a function of photon energy. It shows a strong resonant behavior around 1.68 eV, which corresponds to the lowest-energy exciton peak of $WSe_2$ (Fig. S8b). The polarization rotation decreases sharply when off-resonance on both sides. Based on these results, we conclude that the observed doping profile of $\theta$ is not affected by the optical resonance change. It originates intrinsically from the stripe orders in $WSe_2/WS_2$ moiré superlattice.

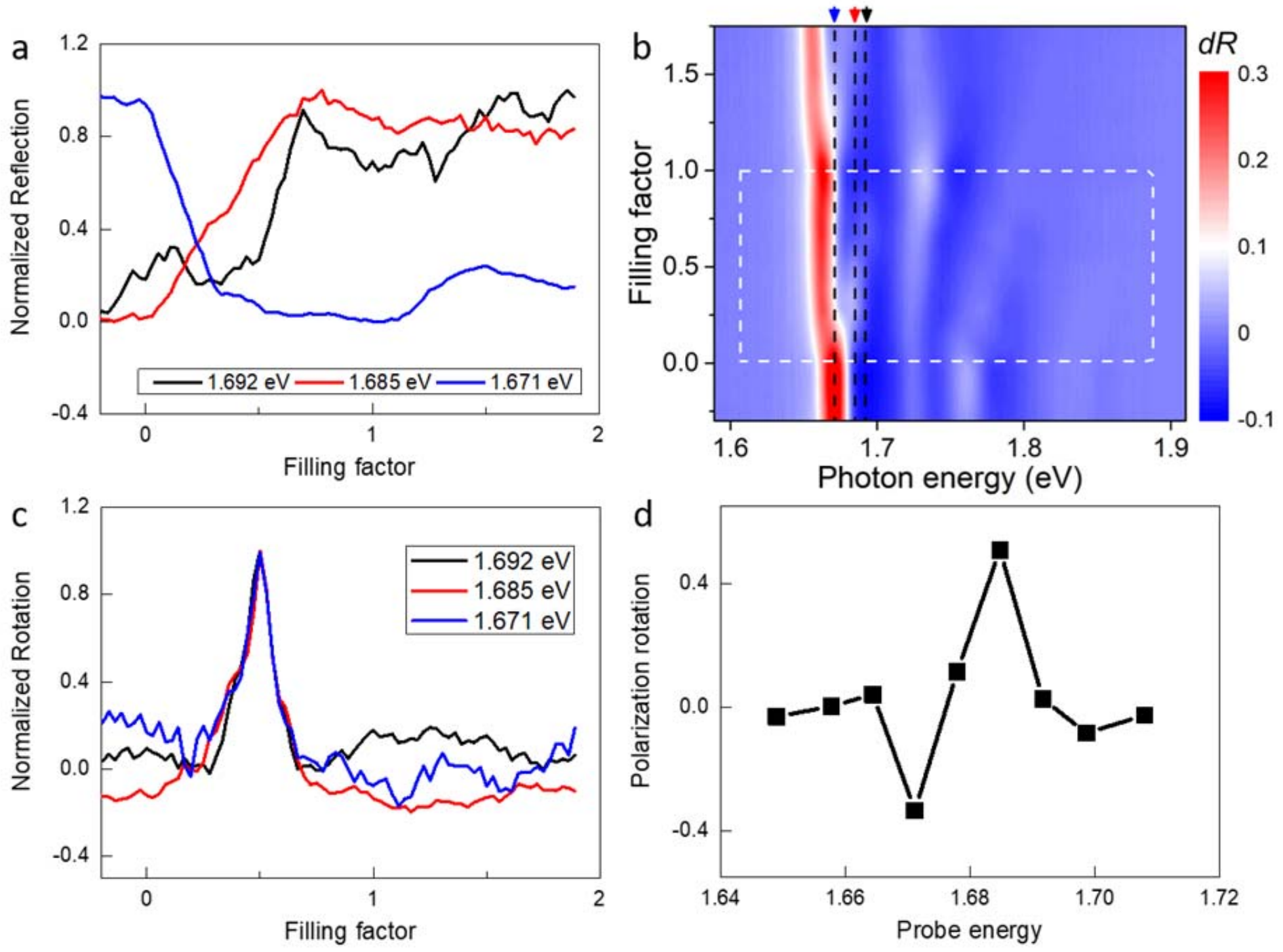

Figure S8. Probe photon energy dependence. **a**, Gate-dependent optical reflection without analyzer P2 at three representative probe energies, labeled by three arrows and vertical dashed lines in (**b**). The reflection change is normalized to [0,1] for each probe energy. **c**, Gate-dependent anisotropy signal measured under the same experimental configuration as (**a**). The maximum absolute rotation is normalized to 1 for each probe energy. The anisotropy profile does not change with probe energy. **d**, Probe energy-dependent anisotropy signal at $\nu = 1/2$ shows a prominent resonance behavior around the lowest-energy exciton energy.

## 8. Thermal cycle effects on stripe domain patterns

Figure S9 shows spatial maps of $\theta$ at $\nu = 1/2$ for device D1 after each thermal cycle for multiple cycles. During each thermal cycle, the sample was first warmed up to room temperature and then cooled down back to 5 K. All other experimental conditions were kept the same. The sample orientation was fixed at $\alpha = 72°$. The stripe domain pattern shows considerable changes after each thermal cycle, suggesting changes in the anisotropy axes.

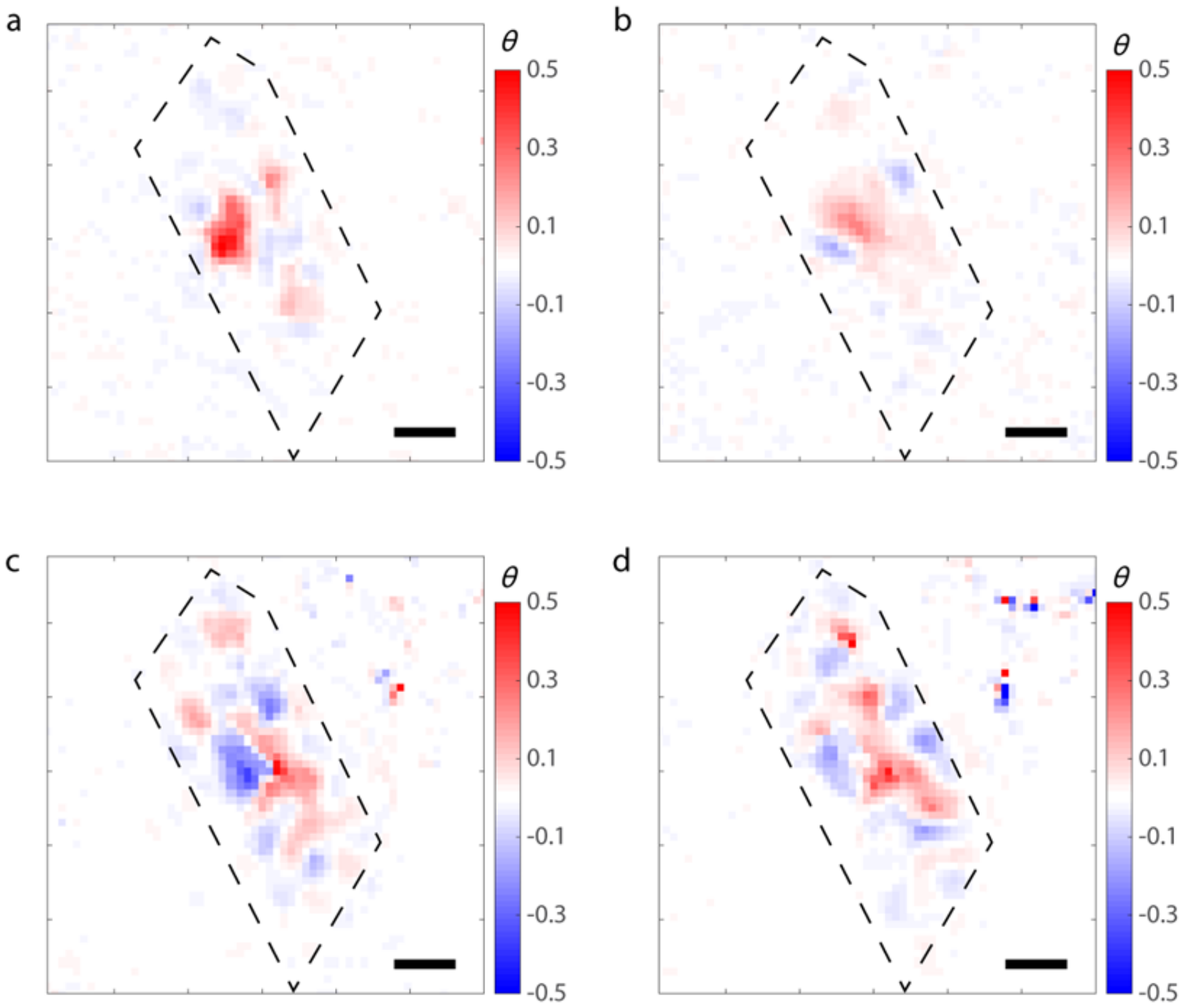

Figure S9. Thermal cycle effects on stripe domains. **a-d**, Spatial maps of $\theta$ at $\nu = 1/2$ in four different thermal cycles. Scale bars represent 2 μm.

To further investigate the thermal cycle effects on anisotropy orientation, we have carried out experiments and analyses the same as Fig. 4 in the main text for another thermal cycle (Fig. S10). Figure 4d and 4e of the main text are also copied here for comparison. Despite the substantial domain pattern change between the two thermal cycles (Fig S10c and S10d), the anisotropy axes remain to preferentially align with the high symmetry directions of the superlattice (Fig S10a and S10b). These observations are consistent with the proposed charge configurations in the main text. The biased alignment to ~ 30 and 120 degree angles (the two distribution peaks) for both thermal cycles indicate the presence of unintentional strain in the sample, which favors alignment to these two particular major axes than the other axes. The disappearance of anisotropy signals at some locations of the sample for the second thermal cycle is likely caused by the formation of stripe domains substantially smaller than the diffraction limit for optical imaging.

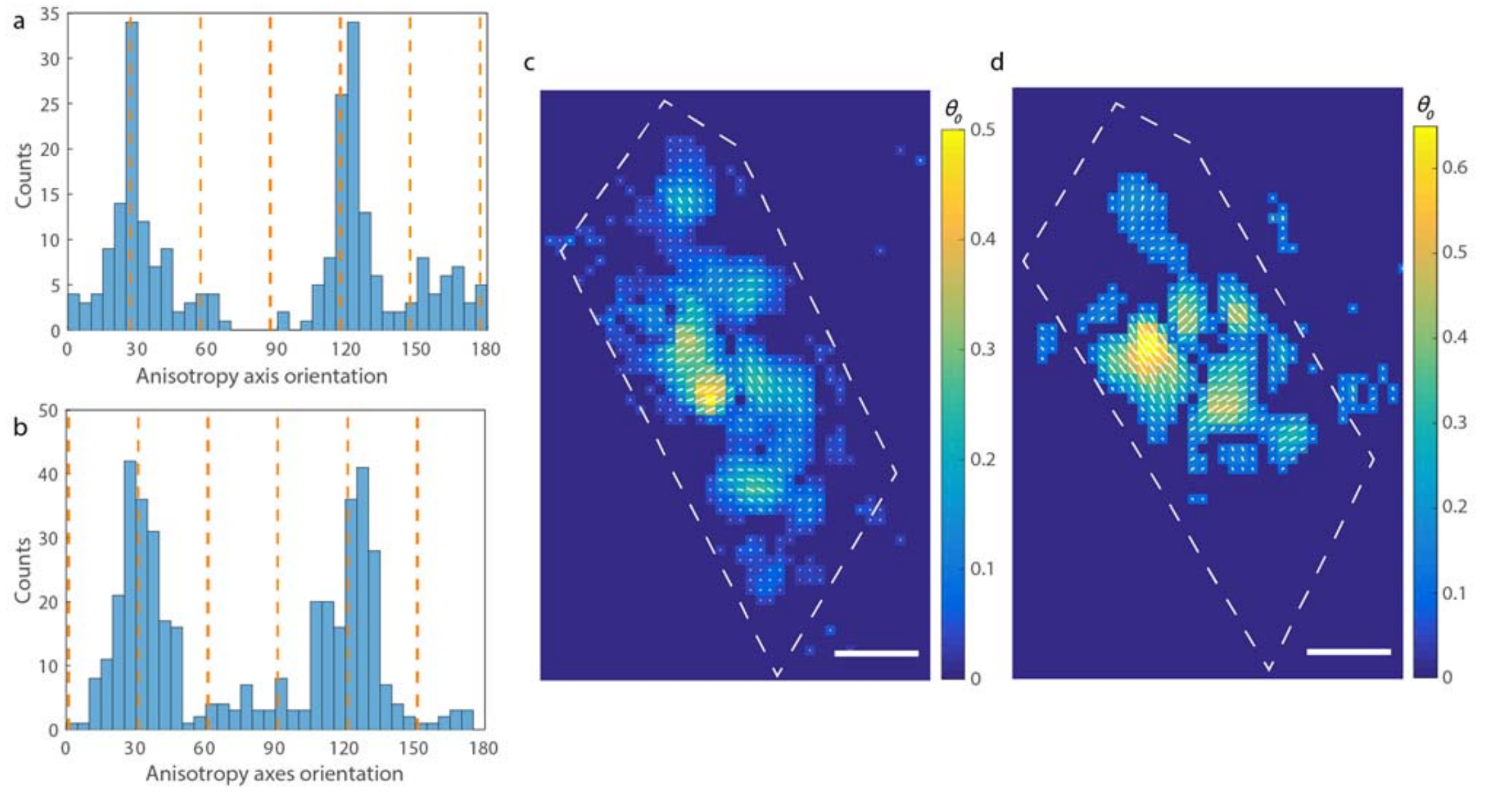

Figure S10. Thermal cycle effects on anisotropy axis. **a-d**, Spatial map of anisotropy amplitude and orientation (**d**) and the anisotropy axis frequency distribution (**b**) for another thermal cycle. Results in main text Fig. 4 is coped as (**a**) and (**c**) for comparison. Scale bars represent 2 μm. The device is mounted in a slightly rotated angle between the two thermal cycles. Therefore, both the heterostructure contour and the high symmetry directions are rotated by 4 degree.

## 9. Possible charge configurations at other commensurate fillings

Figure S11 shows possible charge configurations at fillings of 1/4 (left), 2/5 (middle) and 3/5 (right). The three-fold rotational symmetry is explicitly broken in all cases, consistent with our observation of enhanced electronic anisotropy. The $\nu = 2/5$ state can be viewed as packed domains of stripes at $\nu = 1/2$ and isotropic electron crystal at $\nu = 1/3$. Other possible charge configurations can be found in Ref. [36].

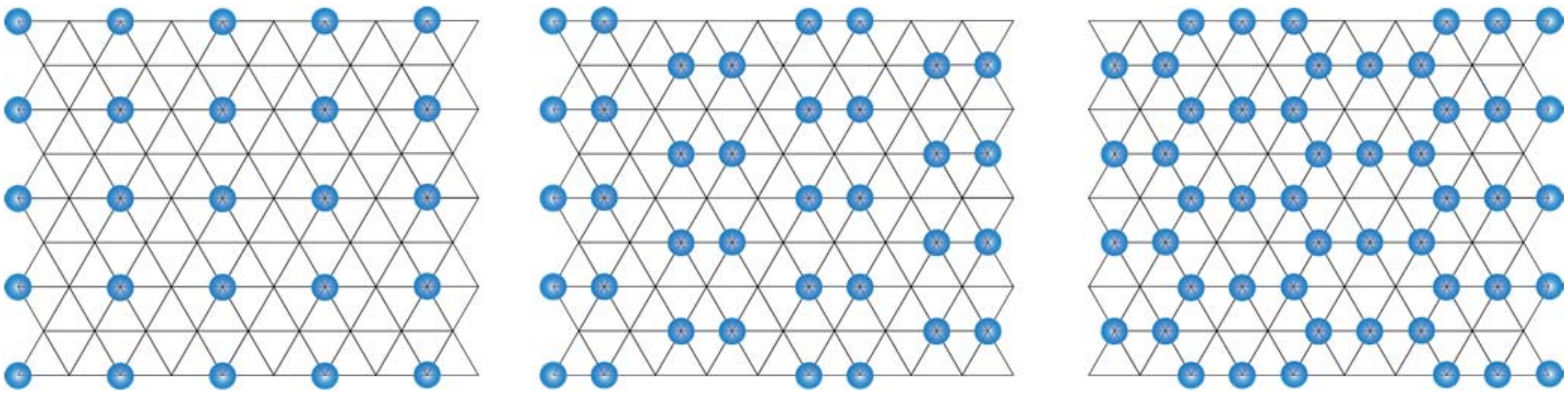

Figure S11. Possible charge configurations at fillings of 1/4 (left), 2/5 (middle) and 3/5 (right).